\documentclass[twocolumn,prl,floatfix]{revtex4}
\usepackage{color}
\usepackage{epsfig}
\usepackage{graphicx}
\usepackage{amsmath}

\newcommand {\be}  {
\begin{equation}
}
\newcommand {\ee}  {
\end{equation}
}
\newcommand {\bea} {
\begin{eqnarray}
}
\newcommand {\eea} {
\end{eqnarray}
}

\newcommand {\ma} {
\mathcal A
}

\newcommand {\mb} {
\mathcal B
}

\newcommand {\mc} {
 \mathcal C
}

\newcommand{\boldnabla}{\text{\boldmath$\nabla$}}

\begin{document}




\title{Achieving realistic interface kinetics in phase field models with a diffusional contrast}
\author{ G. Boussinot$^{1,2}$,
Efim A. Brener$^1$}
\affiliation{$^1$Peter Gr{\"u}nberg Institut, Forschungszentrum J{\"u}lich, D-52425 J{\"u}lich, Germany \\
$^2$Computational Materials Design Department, 
Max-Planck Institut f\"ur Eisenforschung, D-40074 D\"usseldorf, Germany}

\begin{abstract}

Phase field models are powerful tools to tackle free boundary problems. For phase transformations involving diffusion, the evolution of the non conserved phase field is coupled to the evolution of the conserved diffusion field. Introducing the kinetic cross coupling between these two fields [E. A. Brener, G. Boussinot, Phys. Rev. E {\bf 86}, 060601(R) (2012)], we solve the long-standing problem of a realistic description of interface kinetics when a diffusional contrast between the phases is taken into account. Using the case of the solidification of a pure substance, we show how to eliminate the temperature jump at the interface and to recover full equilibrium boundary conditions. We confirm our results by  numerical simulations.


\end{abstract}

\maketitle

{\it Introduction.} 
The phase field method is one of the most important and powerful tool for modeling interface dynamics and pattern formation processes. In particular, it has proven its efficiency for the description of microstructure evolution during phase transformations that are coupled to bulk diffusion. However, anomalous interface kinetic effects exist when a contrast of diffusion coefficient between the phases is taken into account \cite{karma_rappel, almgren, antitrapping1, antitrapping2}. Eliminating them remains a long-standing problem for a realistic phase field modeling.


The phase field method tackles free-boundary problems  with the use of fields that are continuous  across the interface. They exhibit an intrinsic length scale, the interface width $W$, that has to be much smaller than the length scales of the emerging pattern. 
For computationally tractable simulations, $W$ is usually chosen significantly larger than the actual width of the physical interface. However, this leads to an artificial enhancement of the interfacial kinetic effects (for an extended discussion, see Refs. \cite{karma_rappel, almgren,antitrapping1, antitrapping2}). Therefore it is important to eliminate these artificial effects by a proper adjustment of the model parameters. 
For example, for the solidification of a pure substance where the growth is coupled to the heat transport, one may look for an elimination of the jump of temperature at the interface (Kapitza jump). This is the usual approximation in thermal problems. A more particular goal is to recover full equilibrium boundary conditions (no kinetic effects), a situation that corresponds to the limit of small driving forces and that is often met in experiments. 

Classical phase field models do not contain a kinetic cross coupling between the non conserved phase field and the conserved diffusion field, and in this respect are diagonal (model C in the Hohenberg-Halperin nomenclature \cite{hohenberg}). Karma and Rappel developed the so-called thin interface analysis to connect a phase field model and the corresponding macroscopic description \cite{karma_rappel}.
They found that if the diffusivity does not depend on the phase field, i.e. is the same in the two phases, the Kapitza jump is automatically eliminated. They calculated the growth kinetic coefficient and gave the relation between the parameters of the phase field model to recover equilibrium boundary conditions.
However, Almgren showed that, in the presence of a diffusional contrast, the Kapitza jump cannot be eliminated without producing anomalous corrections to the energy conservation law  at the interface \cite{almgren}. 
For isothermal phase transformations in alloys, the same problem exists concerning the jump of diffusion chemical potential (relevant to the solute trapping effect). For the solidification of alloys where the diffusion flux is neglected in the solid (one-sided model), the so-called anti-trapping model was developed to tackle this problem \cite{antitrapping1, antitrapping2}. 
In Ref. \cite{ohno}, an attempt was made to extend this model to a finite diffusion in the solid. Although, in this case, the diffusional fluxes on the two sides of the interface and the interface velocity are related by only one equation (the conservation law), the authors used an additional relation between them (Eq. (5.1) in Ref. \cite{ohno}). Thus, they imposed an unjustified linear relation between two in fact linearly independent quantities \cite{caroli}.

Recently, a kinetic cross coupling between the non-conserved phase field and the conserved diffusion field  was introduced in the phase field equations of motion \cite{onsager, fang, mbe}. 
With this cross coupling, in general allowed by Onsager symmetry, the phase field model becomes non diagonal and possesses an additional kinetic velocity scale compared to classical diagonal models. 
This provides a full correspondence between the number of independent kinetic parameters in the phase field and the macroscopic descriptions.
It was shown that the cross coupling has a crucial importance for some kinetic effects, such as solute trapping in alloys \cite{onsager} and Ehrlich-Schwoebel effect in molecular-beam epitaxy \cite{mbe}. 
Moreover, it was also successfully adapted to the case of the one-sided model \cite{mbe,fang} recovering a full thermodynamical consistency of the anti-trapping model that originally does not obey Onsager symmetry.

In this Letter, we use a non diagonal phase field model \cite{onsager,mbe} to address the long-standing problem of achieving realistic simulations in the presence of a diffusional contrast. In particular, we show that the introduction of the kinetic cross coupling allows to eliminate the Kapitza jump and, in addition, to recover full equilibrium boundary conditions at the interface. This provides a generalization of the results by Karma and Rappel \cite{karma_rappel} in the presence of a diffusional contrast. We test numerically these results using a simple but illustrative example. 

{\it Variational phase field model with kinetic cross coupling.} Complementary to our previous studies on alloys \cite{onsager, mbe}, we focus here on the solidification of a pure material where the phase transformation is coupled to the heat transport. 
We measure all energy densities in units of the latent heat of the transformation $L = T_M(S_L-S_S)$ where $T_M$ is the melting temperature and $S_L$ ($S_S$) is the entropy of the liquid (solid) phase at $T_M$.
We start from an entropy functional $S$:
\be
S = \int_V dV \Big\{ s[e(T,\phi), \phi] - H \left[ (1-\phi^2)^2/4 + (W \boldnabla \phi)^2/2 \right] \Big\}
\ee
where $e$ is the dimensionless internal energy density that depends on the temperature $T$ and on the phase field $\phi$. For simplicity we assume that the cost of an interface is provided by the variations of the phase field only and is described by $H$. The entropy production is
\bea \label{sdot}
\dot S = \int_V dV \left[ \frac{\delta S}{\delta \phi} \dot \phi + \frac{\delta S}{\delta e} \dot e  \right] = \int_V dV \left[ \frac{\delta S}{\delta \phi} \dot \phi + \boldnabla \frac{\delta S}{\delta e} \cdot {\bf J}  \right]
\eea
 where the flux of dimensionless energy $\bf J$ enters the continuity equation
\bea
\dot e = - \boldnabla \cdot {\bf J}. \label{continuity}
\eea
Then the variational equations of motion relate $\delta S/\delta \phi$ and $\boldnabla (\delta S/\delta e)$ to $\dot \phi$ and $\bf J$ through the Onsager linear relations: 
\bea
\frac{c_P T_M^2}{L^2} \; \frac{\delta S}{\delta \phi} &=& \tau \dot \phi + (MW\boldnabla \phi) \cdot {\bf J}, \label{ds_dphi} \\
\frac{c_P T_M^2}{L^2} \; \boldnabla \frac{\delta S}{\delta e} &=&  (MW\boldnabla \phi) \dot \phi +  \frac{\bf J}{D(\phi)} \label{ds_de}, 
\eea 
where $c_P$ is the specific heat at $T=T_M$ that is assumed to be independent of the phase (independent of $\phi$). For simplicity, we consider constant $\tau$ and $M$. The thermal diffusivity $D(\phi)$ depends on $\phi$ to account for a diffusional contrast. While the variational formulations without cross coupling ($M=0$) are well-known in the literature for a long time (see for example Ref. \cite{bi_sekerka} and references therein), cross terms were introduced only recently \cite{onsager,fang,mbe}. They obey Onsager symmetry and are naturally written proportional to $W \boldnabla \phi$. This vector is normal to the interface and is of magnitude of order unity in the interface region and zero in the bulk. It is also important to make a proper link between scalar and vectorial quantities (we refer to Refs. \cite{onsager,mbe} for more details).  A positive entropy production $\dot S>0$ requires 
\be
\tau>0, D(\phi)>0 ,  \Delta = 1-(MW\boldnabla \phi)^2 D(\phi)/\tau >0.
\ee 
 
The variational derivatives $\delta S/\delta \phi = [\partial s/\partial \phi]_e + H[\phi(1-\phi^2) + W^2 \boldnabla^2 \phi]$ and $\delta S/\delta e = [\partial s/\partial e]_\phi$ are evaluated using the thermodynamical relations $[\partial s/\partial \phi]_e = - (L/T) [\partial f(T,\phi)/\partial \phi]_T$ and $[\partial s/\partial e]_\phi = L/T$  \cite{bi_sekerka}. The dimensionless free energy $f=e-Ts/L$ is chosen equal to zero at $T=T_M$. We interpolate the entropies at $T_M$ as:
\be
\frac{T_M}{L} \; s(T=T_M, \phi) = \sigma(\phi) = \frac{\sigma_S + \sigma_L}{2} - \frac{\sigma_L - \sigma_S}{2} \; p(\phi)
\ee 
where $\sigma_S = T_M S_S/L$, $\sigma_L = T_M S_L/L$ and hence $\sigma_L-\sigma_S=1$. 
To be explicit, we choose the switching function $p(\phi)= 15(\phi-2\phi^3/3+\phi^5/5)/8$ that has the properties $p(\phi = \pm 1)=\pm 1$, $p(\phi) = -p(-\phi)$ and $p'(\phi = \pm 1) = p''(\phi = \pm 1)=0$. 
Then,  near $T=T_M$, the dimensionless internal energy $e$ and free energy $f$ read
\be
e(T,\phi) = u + \sigma(\phi) \;\;\; ; \;\;\; f(T,\phi) = -\frac{L}{c_P T_M}  \sigma(\phi) u
\ee 
with the dimensionless temperature $u =c_P(T-T_M)/L$.
Finally, using the above considerations, we present Eqs. (\ref{continuity})-(\ref{ds_de}) in the form of coupled equations of motion for $\phi$ and $u$:
\bea
\Delta \tau \dot \phi = \tilde H \left[ \phi(1-\phi^2) + W^2 \boldnabla^2 \phi \right] - \frac{p'(\phi)}{2} \; u \nonumber \\
 + M W D(\phi) \boldnabla \phi \cdot \boldnabla u  
\eea
and
\be\label{dot_u}
\dot u = \boldnabla \cdot \left \{ D(\phi) \big[ \boldnabla u + M W \dot \phi \boldnabla \phi \big]  \right\} +  \frac{p'(\phi)}{2} \; \dot \phi,
\ee
where $\tilde H = (c_P T_M^2/L^2)H$ is usually a large parameter.

At equilibrium where $u=0$ and $\dot \phi=0$, the phase field obeys $\phi_{eq}(x) = - \phi_{eq}(-x)= -\tanh(x/W\sqrt{2})$ for a solid phase at $x=-\infty$ and a liquid phase at $x=+\infty$, and $W^2[\phi_{eq}'(x)]^2= [1-\phi_{eq}^2(x)]^2/2$. We also define $\sigma_{eq}(x) = \sigma[\phi_{eq}(x)]=(\sigma_S+\sigma_L)/2 - p[\phi_{eq}(x)]/2$. 

{\it Macroscopic description and thin interface limit.} In the macroscopic description, one discusses the thermal diffusion equation in the bulk with boundary conditions at the sharp interface. First, the energy conservation at the interface reads 
\bea
D_S  \boldnabla u|_S \cdot {\bf n}+ V\sigma_S =D_L  \boldnabla u|_L \cdot {\bf n}+V\sigma_L = J_E, \label{temkin}
\eea 
where $D_S$ ($D_L$) is the thermal diffusivity in the solid (liquid) phase, $\bf n$ is the vector normal to the interface, and $\boldnabla u|_S$ ($\boldnabla u|_L$) is the gradient of $u$ on the solid (liquid) side of the interface. $V$ is the normal velocity of the interface and $J_E$ is the normal flux of dimensionless energy through the interface.
Secondly, the temperature at the interface is $T_S$ ($T_L$) on the solid (liquid) side and deviates from $T_M$ due to interfacial kinetic effects. The kinetic boundary conditions relate,  through linear Onsager relations, the difference of chemical potential or free energy across the interface  $\delta f$ and the Kapitza temperature jump $\delta u$ to their conjugate fluxes $V$ and $J_E$. These relations read \cite{balibar, brener_temkin}:
\bea
\delta f = \sigma_S u_S - \sigma_L u_L = \bar \ma V + \bar \mb J_E, \label{def_kinetic1}\\
\delta u = u_L-u_S= \bar \mb V + \bar \mc J_E, \label{def_kinetic2}
\eea
where $u_S=c_P(T_S-T_M)/L$ and $u_L=c_P(T_L-T_M)/L$. 
For a curved interface, $\delta f$ also contains an equilibrium Gibbs-Thomson correction.
$\bar \ma$ is the inverse growth kinetic coefficient (usually denoted by $\beta$ when the other kinetic coefficients are absent), $\bar \mc$ is the Kapitza resistance, and $\bar \mb$ is the cross coefficient. 
A physically motivated procedure for the thin interface limit that involves the entropy production (dissipation function) in the interface region was introduced in Ref. \cite{mbe}. It corresponds to the more mathematically oriented asymptotic matching \cite{karma_rappel, almgren} at its first order. It allows  to express the macroscopic kinetic coefficients $\bar \ma, \bar \mb, \bar \mc$ in terms of the parameters of the phase field model $\tau, M$ and the thermal diffusivity $D(\phi)$ (compare with Eqs. (19)-(21) in Ref. \cite{mbe}): 
\bea
\bar \ma &=& \int_{-\infty}^\infty dx \; [\tau - 2MW \sigma_{eq}(x)] [\phi_{eq}'(x)]^2 \nonumber \\
&&+ \int_{-\infty}^\infty dx \; \left[ \frac{\sigma^2_{eq}(x)}{ D[\phi_{eq}(x)]} - \frac{\sigma_S^2}{2D_S} - \frac{\sigma_L^2}{2D_L} \right], \\
\bar \mb &=& \int_{-\infty}^\infty dx \; MW [\phi_{eq}'(x)]^2 \nonumber \\
&&- \int_{-\infty}^\infty dx \; \left[ \frac{\sigma_{eq}(x)}{ D[\phi_{eq}(x)]} - \frac{\sigma_S}{2D_S} - \frac{\sigma_L}{2D_L} \right], \\
\bar \mc &=& \int_{-\infty}^\infty dx \; \left[ \frac{1}{ D[\phi_{eq}(x)]} - \frac{1}{2D_S} - \frac{1}{2D_L} \right].
\eea
As we see, there is here a larger flexibility for tuning kinetic coefficients compared to classical phase field models where $M=0$.

{\it Elimination of the Kapitza jump and full equilibrium boundary conditions.} The models for the solidification of a pure material usually assume $\delta u=0$. Since $V$ and $J_E$ are linearly independent fluxes, this requires $\bar \mb=\bar \mc=0$. For this purpose we use
\be
\frac{1}{D(\phi)} = \left( \frac{1}{2D_S} + \frac{1}{2D_L} \right) + g(\phi) \left( \frac{1}{2D_S} - \frac{1}{2D_L} \right) \label{d_of_phi}
\ee
where $g(\phi)$ obeys $g(\phi = \pm 1) = \pm 1$ and $g(\phi)=-g(-\phi)$. This choice automatically eliminates the Kapitza resistance $\bar \mc=0$ \cite{footnote}. Then 
\be
\bar \mb = \alpha M - \frac{\gamma W}{2} \left( \frac{1}{2D_S} - \frac{1}{2D_L} \right),
\ee 
where
\bea
\alpha &=& W \int_{-\infty}^\infty dx [\phi_{eq}'(x)]^2 = \frac{2 \sqrt{2}}{3} , \\
\gamma &=&  \int_{-\infty}^\infty \frac{dx}{W} \big \{ 1-p[\phi_{eq}(x)]g[\phi_{eq}(x)] \big \}.
\eea
The condition $\bar \mb=0$ is therefore provided by 
\be\label{mstar}
M=M^* = \frac{\gamma W}{2\alpha} \left( \frac{1}{2D_S} - \frac{1}{2D_L} \right).
\ee
We understand at this point why the kinetic cross coupling was not needed ($M^*=0$) for the Kapitza jump to be eliminated when $D(\phi)=D_S=D_L$ \cite{karma_rappel}.

Using these choices of $D(\phi)$ and $M$, we find the remaining kinetic coefficient $\bar \ma$:
\be
\bar \ma = \frac{\alpha \tau}{W} -\frac{\beta W }{4} \left( \frac{1}{2D_S} + \frac{1}{2D_L} \right)  \label{a}
\ee
where 
\bea
\beta =  \int_{-\infty}^\infty \frac{dx}{W} \big \{ 1-p^2[\phi_{eq}(x)] \big \} \simeq 1.40748 .
\eea
Finally, kinetic effects may be fully eliminated ($\bar \ma=\bar \mb=\bar \mc=0$) by the choice
\be\label{taustar}
\tau=\tau^* = \frac{\beta W^2 }{4\alpha} \left( \frac{1}{2D_S} + \frac{1}{2D_L} \right).
\ee

The condition of stability $\Delta >0$ sets an upper bound for the diffusional contrast if one requires equilibrium boundary conditions. Using the relation $(W\phi'_{eq})^2 = (1-\phi_{eq}^2)^2/2$ and defining the contrast $\nu=(D_L-D_S)/(D_L+D_S)$, the inequality $\Delta = 1-(M^*W\boldnabla \phi)^2 D(\phi)/\tau^* >0$ reads close to equilibrium
\be
1 < \frac{2\beta \alpha}{\gamma^2 \nu^2} \;  \text{ min } \frac{1+\nu g(\phi_{eq})}{(1-\phi_{eq}^2)^2}.
\ee
This means $|\nu| < \nu_{max}$ where $\nu_{max}$ is model dependent through $g(\phi)$ and the coefficient $\gamma$. 
In the case of a larger contrast ($D_S \ll D_L$) which is important for the solidification of alloys, one may use a one-sided model (see the relevant discussion in Ref. \cite{mbe}). 

Two remarks are in order.
First, Almgren has shown that the correction to the conservation law due to the interface stretching effect is eliminated if $\int_{-\infty}^\infty dx [\sigma_{eq}(x) - \sigma_S/2 - \sigma_L/2]=0$ \cite{almgren}. This integral represents the equilibrium interface "adsorption" and vanishes as soon as $p(\phi)=-p(-\phi)$ which is the usual choice in phase field models.
In Ref. \cite{mc_fadden}, the authors effectively add an even part to $p(\phi)$ in order to tune the kinetic coefficient $\bar \mb$ thus allowing for some interface adsorption.
However, it is not physically sound to tune the kinetic effects using a thermodynamical degree of freedom of the model.
Here, we tune the kinetic cross coefficient $\bar \mb$ with the help of the cross coupling term $M$. 
Secondly, surface diffusion is another effect that alters the conservation law at the interface. In order to remove this effect, $D(\phi)$ should satisfy: $\int_{-\infty}^\infty dx [D(\phi_{eq}) - D_S/2 - D_L/2] = 0$ \cite{almgren}. The switching function should in this case be written $g(\phi)=g_0(\phi)+ag_1(\phi)$ where $g_0(\pm 1)=\pm 1$ and $g_1(\pm 1)=0$, both being odd in order to recover $\bar \mc=0$. Then the surface diffusion effect is eliminated with an appropriate choice $a=a^*(|\nu|)$ \cite{almgren}. 

We now shortly summarize how the non diagonal phase field model is designed. With the switching function $p(\phi)$, one tunes the thermodynamic adsorption property of the interface that causes a kinetic correction to the conservation law due to interface stretching. With the diffusivity $D(\phi)$, one tunes the kinetic coefficient $\bar \mc$ and the kinetic correction to the conservation law due to surface diffusion. With the cross coupling coefficient $M$, one tunes the kinetic coefficient $\bar \mb$. With the relaxation time $\tau$, one tunes the kinetic coefficient $\bar \ma$. Here we have presented the procedure to  eliminate {\it simultaneously} all the kinetic effects. Let us note that an interface anisotropy may be introduced in a routine way through the orientation dependences $W({\bf n}), \tau({\bf n})$ \cite{karma_rappel} and $M({\bf n})$.

{\it Numerical test.} We now perform a numerical validation of the presented theory using a simple but illustrative example. We discuss the solidification of a thin film that may exchange heat with an environment at temperature $T_E < T_M$. Averaging the temperature in the film over its thickness the problem becomes two-dimensional. Moreover, we discuss the propagation of a flat solidification front and the average dimensionless temperature in the film then obeys the one-dimensional diffusion equation: 
\be
\dot u(x) = D_k  u''(x) -\frac{u(x)-u_E}{\tau_v}
\ee 
in phase $k=L, S$. We model the heat exchange with the environment using a simple linear law valid for small temperature differences between the film and the environment, i.e. for $-u_E = -c_P(T_E-T_M)/L \ll 1$. Here $\tau_v$ is the characteristic time for this exchange.

We focus on the case without Kapitza jump, $\bar \mb=\bar \mc=0$, that requires $M=M^*$. Then, according to Eqs. (\ref{def_kinetic1}), (\ref{def_kinetic2}), and due to the fact that $\sigma_L-\sigma_S=1$, we have $u_S=u_L=u_i=-\bar \ma V$ where $\bar \ma$ is given by Eq. (\ref{a}). 

 In a co-moving frame of reference with the interface located at the origin $x=0$, phase $S$ occupying $x<0$ and phase $L$ occupying $x>0$, the solution in the quasi-static approximation ($\dot u=0$) reads
\bea
u(x<0) = u_E + (u_i-u_E) \exp \left(x \big/\sqrt{D_S \tau_v} \right), \nonumber \\
u(x>0)= u_E + (u_i-u_E) \exp \left(-x \big/\sqrt{D_L \tau_v} \right). \label{u(x)}
\eea
Using the energy conservation at the interface, i.e. $V=D_S u'(x=0^-) - D_L u'(x=0^+)$, we find
\bea
V = V_{eq} \left( 1- \frac{ \bar \ma (J_S+J_L)}{1+\bar \ma (J_S+J_L)}\right) \label{velocity} 
\eea
where $J_S = \sqrt{D_S/\tau_v}$ and $J_L=\sqrt{D_L/\tau_v}$, and with
\be\label{veq}
V_{eq} = -u_E(J_S+J_L)
\ee
being the velocity for equilibrium boundary conditions, i.e. when $u_i=0$.  Let us note that the quasi-static approximation that allows one to use Eqs. (\ref{u(x)}) is justified for $-u_E \ll 1$.

We perform phase field simulations  where the temperature evolution equation (\ref{dot_u}) is replaced by
\be\label{dot_u_mbe}
\dot u = \boldnabla \cdot \left\{ D(\phi) \left[ \boldnabla u + MW \dot \phi \boldnabla \phi \right] \right\}+ \frac{p'(\phi)}{2} \; \dot \phi  - \frac{u-u_E}{\tau_v}
\ee
that takes into account the heat exchange with the environment.
We use  $u_E=-0.1$, $D_S \tau_v/W^2 = 200$ and different ratios $D_L/D_S = (1+\nu)/(1-\nu)$. We choose $g(\phi)=\phi$ (no need to eliminate surface diffusion here) for which $\nu_{max} \simeq 0.71$ (for example for the choice $g(\phi)=p(\phi)$, one has $\nu_{max} \simeq 0.86$).
 \begin{figure}[htbp]
\includegraphics[width=\linewidth]{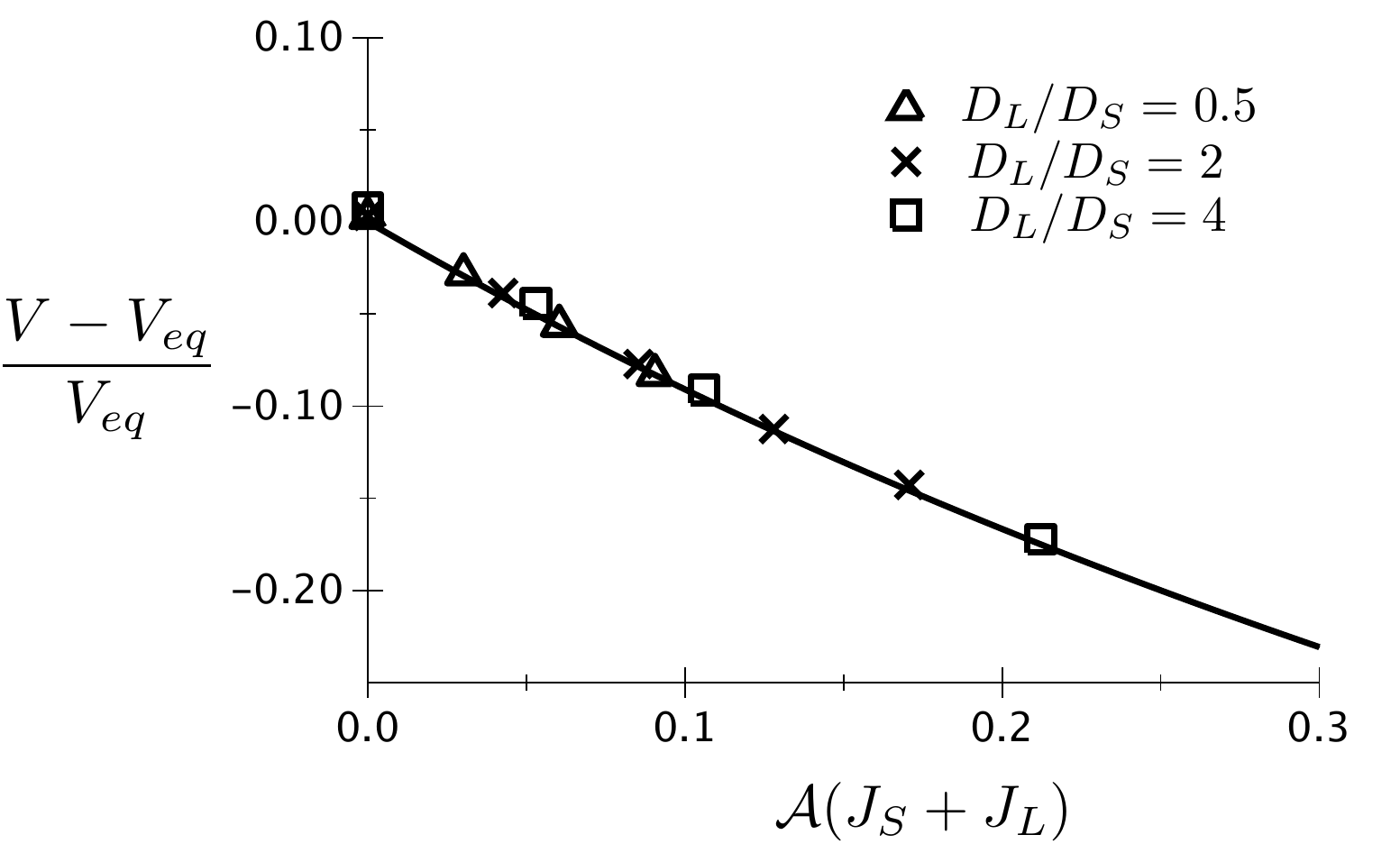}
\caption{\label{b=0} $(V-V_{eq})/V_{eq}$ as a function of $\ma (J_S+J_L)$ for different values of $D_L/D_S$ from simulations (triangles, crosses, squares) and from analytics (line).}
\end{figure}
In Fig. \ref{b=0}, we present $(V-V_{eq})/V_{eq}$ as a function of $\ma(J_S +J_L)$ where $V$ is obtained from the simulations (symbols) and $V_{eq}$ is given by Eq. (\ref{veq}). We compare with the analytical prediction (line) with $V$ given by Eq. (\ref{velocity}). 
The excellent quantitative agreement between simulations and analytics provides a strong support for the theory.

Let us note that this problem is formally close to the problem of a step growth in molecular-beam-epitaxy where the diffusion coefficient differs from one terrace to the other, for example due to a different surface reconstruction \cite{frisch}.


{\it Summary.}  
In this Letter, we have achieved a realistic description of interface kinetics in a phase field model with a diffusional contrast.
This is enabled by a kinetic cross coupling (non diagonal model) between the conserved and the non conserved fields that is not present in classical diagonal models. 
This cross coupling provides a full correspondence between the number of independent kinetic parameters in the phase field and the macroscopic descriptions. This correspondence allows to tune the kinetic effects as desired and for example allows to eliminate them partly or fully.
Using the case of the solidification of a pure substance, we first eliminate the Kapitza temperature jump at the interface and secondly fully eliminate interfacial kinetic effects. 
We obtain an excellent quantitative agreement between phase field simulations and the analytical solution of the corresponding macroscopic approach using a simple but illustrative example. 
The extension of our results to multi-phase systems where triple junctions are present might be challenging.


We are grateful to R. Spatschek for useful discussions. E.A.B. acknowledges the support of the Deutsche Forschungsgemeinschaft under Project SFB 917.


\begin{thebibliography}{99}
\bibitem{footnote} Another goal may be to have a finite $\bar \mc$. Then a possibility is to add to $1/D(\phi)$ an even function of $\phi$ that vanishes in the bulk (see for example Ref. \cite{mbe}).
\end{thebibliography}
\end{document}